\documentclass[acmtog, authorversion]{acmart}
\acmSubmissionID{271}
\newcommand{\jungnam}[1]{\textcolor{blue}{}}
\newcommand{\moon}[1]{\textcolor{green}{}}
\newcommand{\jehee}[1]{\textcolor{magenta}{}}
\newcommand{\jungdam}[1]{\textcolor{orange}{}}
\newcommand{\TODO}[1]{\textcolor{red}{}}

\usepackage{ dsfont }

\usepackage{subcaption}
\usepackage{soul}

\AtBeginDocument{%
  \providecommand\BibTeX{{%
    \normalfont B\kern-0.5em{\scshape i\kern-0.25em b}\kern-0.8em\TeX}}}

\copyrightyear{2023}
\acmYear{2023}
\setcopyright{acmlicensed}\acmConference[SIGGRAPH '23 Conference Proceedings]{Special Interest Group on Computer Graphics and Interactive Techniques Conference Conference Proceedings}{August 6--10, 2023}{Los Angeles, CA, USA}
\acmBooktitle{Special Interest Group on Computer Graphics and Interactive Techniques Conference Conference Proceedings (SIGGRAPH '23 Conference Proceedings), August 6--10, 2023, Los Angeles, CA, USA}
\acmPrice{15.00}
\acmDOI{10.1145/3588432.3591492}
\acmISBN{979-8-4007-0159-7/23/08}

\begin{document}

\title[Bidirectional GaitNet]{Bidirectional GaitNet: A Bidirectional Prediction Model of Human Gait and Anatomical Conditions}

\author{Jungnam Park}
\email{jungnam04@imo.snu.ac.kr}
\affiliation{%
  \department{Department of Computer Science and Engineering}
  \institution{Seoul National University}
  \country{South Korea}
}

\author{Moon Seok Park}
\email{pmsmed@gmail.com}
\affiliation{%
  \department{Department of Orthopaedic Surgery}
  \institution{Seoul National University Bundang Hospital}
  \country{South Korea}}

\author{Jehee Lee}
\email{jeheel@ncsoft.com}
\affiliation{%
  \institution{NCsoft}
  \country{South Korea}}
\affiliation{%
  \department{Department of Computer Science and Engineering}
  \institution{Seoul National University}
  \country{South Korea}}

\author{Jungdam Won} 
\authornote{corresponding author}
\orcid{0000-0001-5510-6425}
\email{jungdam@imo.snu.ac.kr}
\affiliation{%
 \department{Department of Computer Science and Engineering}
 \institution{Seoul National University}
 \country{South Korea}
}


\begin{abstract}
We present a novel generative model, called Bidirectional GaitNet, that learns the relationship between human anatomy and its gait. The simulation model of human anatomy is a comprehensive, full-body, simulation-ready, musculoskeletal model with 304 Hill-type musculotendon units. The Bidirectional GaitNet consists of forward and backward models. The forward model predicts a gait pattern of a person with specific physical conditions, while the backward model estimates the physical conditions of a person when his/her gait pattern is provided. Our simulation-based approach first learns the forward model by distilling the simulation data generated by a state-of-the-art predictive gait simulator and then constructs a Variational Autoencoder (VAE) with the learned forward model as its decoder. Once it is learned its encoder serves as the backward model. We demonstrate our model on a variety of healthy/impaired gaits and validate it in comparison with physical examination data of real patients.
\end{abstract}



\maketitle

\section{Introduction}

Realistic simulation of human movement is one of the long-standing challenges in computer graphics. Musculoskeletal simulation has provided stepping stones to reproduce a range of human movements at the biomechanical and anatomical levels, adding significant realism to the movements created. It can also be used to predict how changes in anatomical conditions (e.g., bone deformity, muscle capacity/deficiency, mass distribution) and intrinsic/extrinsic factors (e.g., metabolic energy expenditure, fatigue, pain) affect human movement.

Accurate setting of anatomical conditions is the very first and fundamental process for generating realistic movements in musculoskeletal simulation as it creates the space for anatomically plausible human movements. A number of biomechanical studies have been conducted to accurately model these conditions through various experiments on human subjects and cadavers~\cite{arnold2010model, rajagopal2016full, delp1990interactive, delp2007opensim, carbone2015tlem}. Several standard models for the typical/average human body have been adopted by the research community~\cite{delp1990interactive, delp2007opensim} and those average body models have significantly contributed to recent progress in biomechanics research~\cite{dembia2020opensim, lee2019scalable, park2022generative}. Building an accurate musculoskeletal model of a specific (healthy or impaired) person has long been a notoriously difficult challenge since accurate modeling of live organs and tissues often requires invasive measurements.

In this paper, we address the problem of estimating the physiological parameters of a musculoskeletal model from observed gait cycles. Our anatomy model describes the conditions of individual bones and muscles with 300+ parameters. Conceptually speaking, we want to uncover the relationship between human anatomy and gait. Although this relationship is well accepted empirically in biomechanics and clinical gait analysis, the relationship is probabilistic rather than a one-to-one deterministic mapping. For example, many people with different physical conditions can walk with a similar gait, and conversely, there is no guarantee that two people with similar physical conditions will walk with a similar gait. In this paper, we build a novel generative model, which we call {\em Bidirectional GaitNet}, based on a comprehensive full-body, simulation-ready, musculoskeletal model. The {\em Bidirectional GaitNet} consists of forward and backward models. The forward model is functionally equivalent to predictive gait simulation, which generates a bipedal gait for any anatomical model with a specific set of physical conditions. Conversely, the backward model is its inverse process that estimates the physical conditions of the model given a gait. More specifically, we first learn the forward model by distilling the simulation data generated by a state-of-the-art predictive gait simulator and then construct a conditional Variational Autoencoder (c-VAE) with the forward model as its decoder. Once it is learned, its encoder serves as the backward model. By the nature of VAE, our backward model generates a distribution of physical conditions that potentially produce the input gait in predictive gait simulation, from which many different physical conditions can be sampled.

We demonstrate the power of our model by showing results with a variety of healthy/impaired gaits. The simulated results are validated in comparison to not only unseen simulated gaits but also gaits from real patients. The effectiveness of the non-trivial system design choices that we made for developing our model are also validated by ablation studies. Code for this paper is available at https://github.com/namjohn10/BidirectionalGaitNet.

\section{Related work}

In musculoskeletal simulation, the human body is typically modeled by rigid bones and flexible musculotendon units. The bones are connected by rotational joints to which the musculotendon units are attached such that they can actuate the joints.  The muscle dynamics is often formulated using Hill-type muscles~\cite{zajac1989muscle, delp2007opensim} composed of contractile and elastic elements and the dynamics of each element is determined by force-length and force-velocity curves. Accurate simulation requires adequate determination of all anatomical conditions of the bones and the muscles based on reliable measurements.  A set of parameters for a typical/average human have been estimated by taking measurements from live tissues and cadavers~\cite{delp1990interactive, rajagopal2016full, arnold2010model, carbone2015tlem}. Estimates thus obtained have been used as default parameters in many simulation-based studies~\cite{delp2007opensim, lee2019scalable, dembia2020opensim}.  
Anthropometric scaling of a musculoskeletal model can generate a range of models with different heights, weights, and limb lengths~\cite{ryu2021functionality}. Building accurate individualized models often requires expensive medical images (e.g., CT and MRI) and labor-intensive image labeling~\cite{matias2009transformation, levin2011extracting, li2022nimble}.

Building robust dynamic controllers that can drive musculoskeletal models has been regarded as an open challenge for decades because musculoskeletal models are high-dimensional and highly nonlinear, and their control systems are often partly under-actuated and, at the same time, partly over-determined. In this paper, we will focus on legged locomotion, although there have been a series of studies for simulating other body parts, such as upper bodies~\cite{lee2009comprehensive, lee2018dexterous}, eyes~\cite{nakada2018deep}, faces~\cite{ichim2017phace, cong2016art}, hands~\cite{sachdeva2015biomechanical} and swimming~\cite{si2014realistic}.

Biomechanics researchers have developed locomotion controllers to answer research questions, such as how weakening certain muscles affect gait patterns.  Open-loop ~\cite{sok2007simulating, liu2010sampling, borno2013trajectory, falisse2019rapid}, model-based feedback control ~\cite{geyer2010muscle, yin2007simbicon, lee2010data, ye2010optimal, coros2010generalized, ha2012falling, liu2016guided, song2015neural, ong2019predicting}, and their combinations~\cite{mordatch2014combining} have been explored where the controller design is often motivated by structural and functional hypotheses on nervous and motor control systems. On the other hand, computer animation researchers have focused on allowing physically-simulated characters to move lifelikely as humans by adding actuation constraints imposed by muscle dynamics. Wang et al.~\shortcite{wang2012optimizing} developed walking controllers for 3D biped characters equipped with 8 hill-type muscles per leg with the muscle-reflex model proposed by Geyer and Herr ~\shortcite{geyer2010muscle}. Stochastic optimization is used to determine feedback control parameters such that biped characters can maintain their balance while walking. Geijtenbeek et al.~\shortcite{geijtenbeek2013flexible} applied a similar approach to non-human musculoskeletal characters, where manually-specified initial muscle routings are further optimized to improve the motor skills of the characters. Lee et al.~\shortcite{lee2014locomotion} demonstrated interactive controllers for full-body musculoskeletal characters having more than 100 muscles, where it first runs offline optimization to refine the reference motion and then the online controller based on quadratic programming tracks the refined reference motion at runtime.  


Recently, deep reinforcement learning (DRL) has successfully demonstrated its capabilities in solving high-dimensional, continuous control problems including human motion imitation~\cite{yu2019figure, peng2018deepmimic, bergamin2019drecon, park2019learning, won2020scalable, peng2021amp, 
Merel:2019:neural, peng2022ase, won2022physics}, motion control in complex 
environments ~\cite{clegg2018learning, liu2018learning, won2021control, yang2022learning, ye2022neural3points, winkler2022questsim} and non-human character control~\cite{yu2018learning, luo2020carl, lee2022learning, ishiwaka2022}. The control of musculoskeletal characters is no exception for these technological innovations; in particular, controllers based on DRL have been significantly improved in terms of robustness against external perturbation, computational efficiency at runtime, and the scope of reproducible motor skills. Many simulation algorithms for controlling biped musculoskeletal models competed in the \textit{Learn-to-Move} challenges~\cite{kidzinski2020artificial} and DRL-based algorithms performed well. The winner of the final challenge proposed a model-based DRL algorithm equipped with an ensemble of probabilistic dynamics models and a risk minimization scheme by expanding the lower confidence bound of the value estimation~\cite{kidzinski2018learning}.  Lee et al.~\shortcite{lee2019scalable} developed a two-level control architecture composed of a trajectory mimicking network and a muscle coordination network, by which comprehensive musculoskeletal characters with more than 300 muscles successfully reproduced highly dynamic human movements such as running, jumping, and cartwheel.  Yifeng et al.~\shortcite{jiang2019synthesis} proposed a new action space for DRL that mimics the behaviors of musculoskeletal simulation computationally more efficiently with torque-based simulation.  Park et al.~\shortcite{park2022generative} presented a predictive gait simulation framework \textit{Generative GaitNet}, which can predict a broad spectrum of healthy and pathological gait of comprehensive musculoskeletal models over a high-dimensional parameter space spanned by anatomical (e.g., bone/muscle parameters, mass distribution, muscle capacity) and gait (e.g., stride and cadence) conditions.

\section{human anatomy and gait}

The musculoskeletal model in this work is designed to represent healthy and pathological gaits often discussed in clinical gait analysis and medical engineering. Our model consists of 23 bones connected by 22 skeletal joints and 304 musculotendons. The activation of muscles generates contraction forces that drive skeletal joints. The individual bones and muscles are conditioned by anatomical parameters. Let $C_\mathrm{anatomy}=(C_\mathrm{skeleton}, C_\mathrm{muscle})$ be the anatomical condition, where $C_\mathrm{skeleton}=(c_\mathrm{head},c_\mathrm{trunk},c_1,\cdots,c_8,\tau_1, \tau_2)$ represents the scaling factors of the head, the trunk, the lengths of four (upper and lower) limbs with respect to a reference (average) model of healthy adults. $\tau_1$ and $\tau_2$ are the torsional angles of femurs, which correspond to femoral anteversion and retroversion often observed in pathological gait. Each musculotendon is conditioned by two parameters: weakness and contracture. The weakness parameter of a muscle indicates the ratio of maximum isometric force the muscle can exert relative to the reference model. The higher the parameter, the larger the force the muscle can produce. The contracture parameter indicates the scaling factor of the muscle length relative to the reference model, which refers to the permanent shortening of a muscle that often limits the range of joint movements.

Conventionally, a complete two-step cycle of gait begins with a left heel strike, where both feet are in contact with the ground at the same time, and ends at the next left heel strike. To avoid the singularity at both ends of the gait cycle, two gait cycles are considered as basic units of gait in our system. Therefore, the gait pattern is represented by a time series of full-body poses $M=(Q_{0}, Q_{\delta}, Q_{2\delta}, \cdots)$, where each pose $Q=(\mathbf{h}, \mathbf{v}, \mathbf{q}_0,\mathbf{q}_1, \mathbf{q}_2,\cdots)$ denotes the root (pelvis) height $h$ from the ground, the root linear velocity $\mathbf{v}$ parallel to the ground plane, and joint rotations $\mathbf{q}_i$ with respect to their parent joints. $\mathbf{q}_0$ denotes the global orientation of the root body node. We use the first two columns of the rotation matrix to describe the rotation. The sample rate $\delta$ is 60 per two gait cycles in our implementation. The gait is conditioned by two parameters: stride and cadence. 
\begin{figure}
    \includegraphics[width=\linewidth]{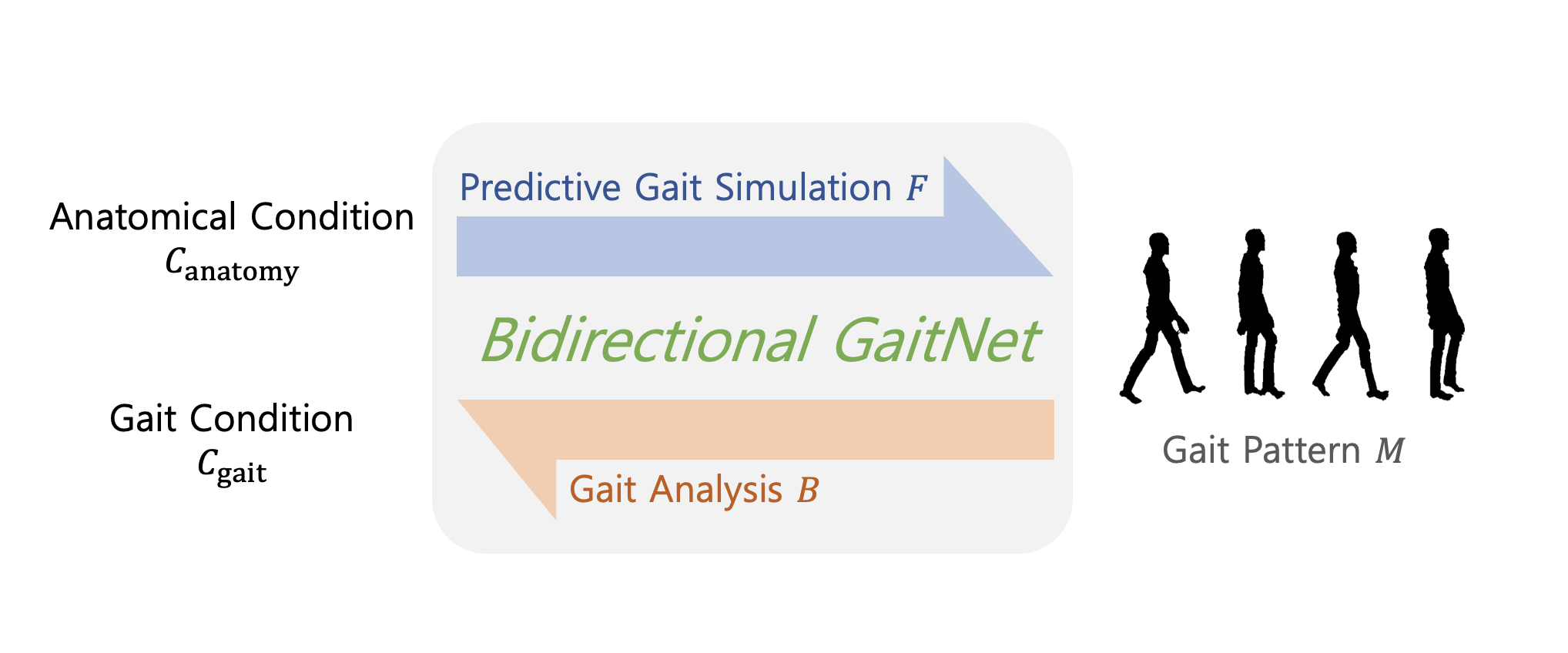}
    \caption{\label{fig:overview} Predictive gait simulation and intelligent gait analysis. }
\end{figure}

The predictive gait simulation generates a gait pattern $M$ that likely occurs given anatomical and gait conditions (see Figure~\ref{fig:overview}). Conversely, gait analysis is its inverse process of predicting the anatomical and gait conditions of a given gait pattern. The human musculoskeletal model is a dynamical system that is partly under-actuated because the root node is not actuated and, at the same time, partly over-actuated because the body has more muscles than minimally required to drive the skeletal joints. This redundancy makes the whole process probabilistic.

The predictive power of gait simulation stems from making full use of the laws of physics, providing accurate anatomical modeling, and designing reliable control policies for seemingly unstable biped locomotion. Recent approaches successfully derived robust control policies through deep reinforcement learning~\cite{lee2019scalable,kidzinski2018learning}. In this case, the core of a predictive gait simulation is a policy network, which takes the current state of the musculoskeletal model as input and outputs the desired level of activation at all muscles. The state-of-the-art method learned control policies conditioned by high-dimensional conditioning vectors (more than 600 parameters)~\cite{park2022generative}. The musculoskeletal model with specific physical conditions driven by a policy network generates a trace of full-body poses in physics-based simulation. 

In this paper, the key challenge is to find the inverse process of predictive simulation. Given a gait pattern, the goal of gait analysis is to find corresponding physical conditions. There are three issues to be addressed. First, the search space is high-dimensional. In our system,  conditioning vectors are 280 dimensional. State space search or optimization in such a high-dimensional space is often computationally intractable. Secondly, the solution is not unique. As discussed before, the backward mapping is probabilistic, and thus we need to find a probabilistic distribution over the conditioning space rather than a single optimal solution. Lastly, the computational cost should be reasonable. Rolling out a single gait pattern from a predictive simulator requires the computational cost of physics-based simulation over a complete gait cycle, which is substantial with a stochastic sampling of gait patterns. 

We address these issues by pretraining forward and backward networks. We transfer control policies from policy networks to a forward network representing a direct anatomy-to-gait mapping. The transfer process is similar to policy distillation. It takes samples from the policy network and learns the target network using supervised regression. The pretrained forward network has two advantages. The forward network generates physically-valid gait patterns without actually performing physics-based simulations because it imitates the behavior of the policy network. This direct anatomy-to-gait mapping is computationally more efficient at runtime than predictive gait simulation. This computational efficiency also makes it computationally feasible to pretrain the backward network employing a conditional Variational AutoEncoder (c-VAE) that models the probabilistic mapping between anatomy and gait using Gaussian distributions in latent space.

\section{Bidirectional GaitNet}
\label{sec:method}

Let $C_\mathrm{anatomy}=(C_\mathrm{skeleton},C_\mathrm{muscle})$ and $C_\mathrm{gait}$ be anatomical and gait conditions, respectively. 
The gait pattern $M=(Q_{0}, Q_{\delta},Q_{2\delta}, \cdots, Q_{59\delta})$ includes two gait cycles, which are parameterized by phase $\phi \in[0,4\pi]$. 
Our \textit{Bidirectional GaitNet} learns the relationship between anatomy and gait by conditional forward mapping $\mathbf{\hat F}(M|C_\mathrm{anatomy}, C_\mathrm{gait})$ and backward mapping $\mathbf{\hat B}(C_\mathrm{anatomy}, C_\mathrm{gait}|M)$. This formulation has several issues in designing network models. The size of the forward network depends on the frame rate $\delta$ and can be larger than actually needed. Given a gait pattern, its stride, cadence, body proportions, and limb lengths are readily available in the motion capture process or can be directly computed from the gait data. To address the issues, we simplify the forward and backward mappings, respectively, by $\mathbf{F}(Q_{\phi}|C_\mathrm{anatomy}, C_\mathrm{gait},\phi)$ and $\mathbf{B}(C_\mathrm{muscle}|M, C_\mathrm{skeleton}, C_\mathrm{gait})$ where we added the phase in the input layer of the forward network and reduced the output layer such that it generates a full-body pose $Q_{\phi}$ at phase $\phi$ instead of a full gait pattern. This design decision significantly reduces the complexity of the prediction and allows us to achieve improved accuracy with smaller regression networks. The backward network also becomes smaller by removing the skeleton and gait conditions from the output layer.

\subsection{Forward GaitNet}

The \textit{Forward GaitNet} is a regression network learned in a supervised manner with training data generated by a predictive gait simulator. In our work, we use Generative GaitNet~\cite{park2022generative} to generate a collection of condition-gait tuples $\{(C^i_\mathrm{anatomy}, C^i_\mathrm{gait}, M^i)\}$ randomly sampled over the domain of anatomy and gait conditions. The loss function for regression is
\begin{equation}
    \Sigma_i \Sigma_\phi \mathrm{D}_{\mathrm{pose}}(Q^i_{\phi}, \mathrm{FGN}(\phi, C^i_\mathrm{anatomy}, C^i_\mathrm{gait})),
\label{eq:loss:FG}
\end{equation}
where $\mathrm{FGN}$ is the output of the regression network. $\mathrm{D}_\mathrm{pose}$ measures the difference between two full-body poses
\begin{equation}
\begin{split}
    \mathrm{D}_{\mathrm{pose}}(Q, Q') = w_h \|\mathbf{h}-\mathbf{h}'\|^2 + w_v \|\mathbf{v}-\mathbf{v}'\|^2 
    + \Sigma_j \|\mathrm{D}_\mathrm{rot}(\mathbf{q},\mathbf{q}')\|^2,
\end{split}
\end{equation}
where it computes the weighted sum of the differences in root heights, root velocities, and joint rotations. $w_h$ and $w_v$ are the weights of their corresponding terms and $\mathrm{D}_\mathrm{rot}$ is the difference between two rotation matrices.

\begin{figure}
    \includegraphics[width=\linewidth]{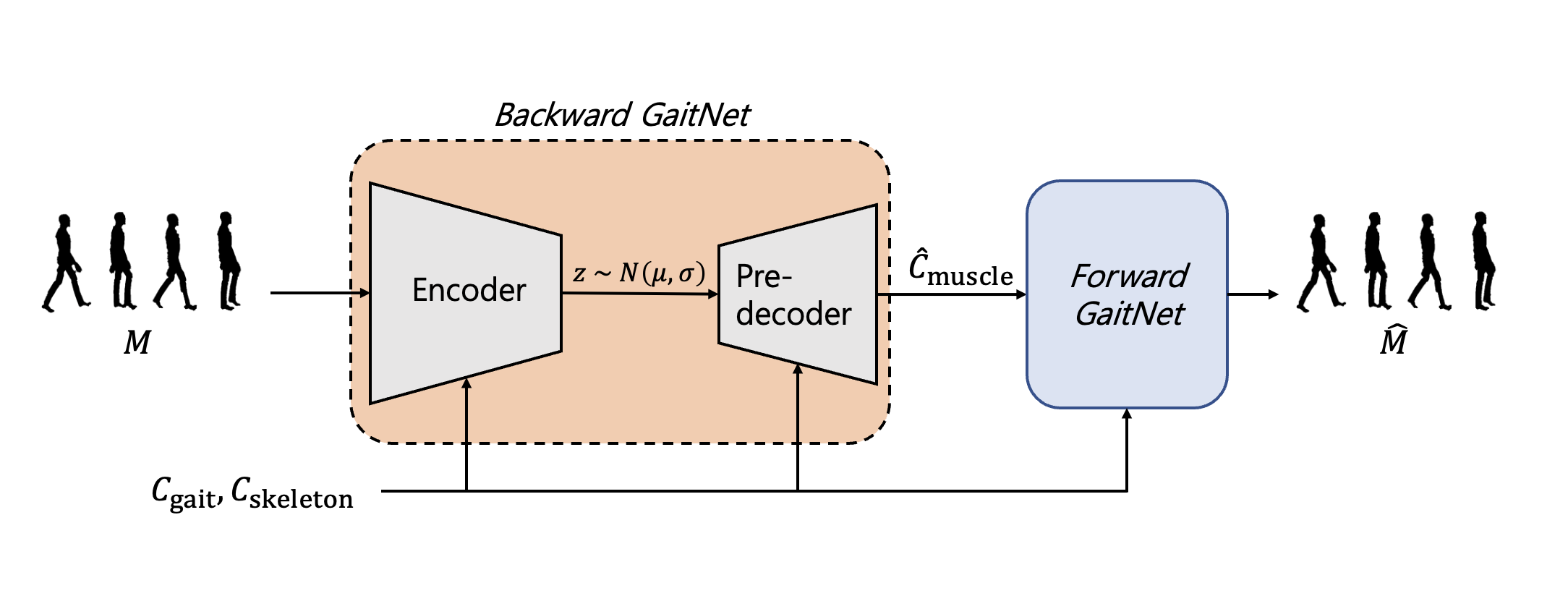}
    \caption{\label{fig:bidirectional_gaitnet} The structure of Bidirectional GaitNet.}
\end{figure}

\subsection{Backward GaitNet}

The \textit{Forward GaitNet} and the \textit{Backward GaitNet} constitute a conditional Variational AutoEncoder (c-VAE) that takes a gait pattern $M$ as input (see Figure~\ref{fig:bidirectional_gaitnet}). Once the \textit{Forward GaitNet} is learned through regression, we learn the \textit{Backward GaitNet} while having the parameters of the \textit{Forward GaitNet} fixed. The loss function is
\begin{equation}
    w_\mathrm{g} \mathrm{D}_\mathrm{g}(M, \hat{M}) + 
    w_\mathrm{kl}\mathrm{D}_\mathrm{kl}(N(\mu , \sigma)||N(0,I)) + \sum_m{w^m(1 - \hat{c}_m)^2},
\label{eq:loss:BG}
\end{equation}
where $\hat M$ is a gait pattern reconstructed through the autoencoder.
\begin{equation}
    \hat{M} = \mathrm{FGN}(\mathrm{BGN}(M,C_\mathrm{gait},C_\mathrm{skeleton}), C_\mathrm{gait},C_\mathrm{skeleton}).
\end{equation}
Note that $\mathrm{FGN}$ here is a gait pattern generated by concatenating the output of the phase-wise forward network over two gait cycles. $D_{\mathrm{kl}}$ measures KL-divergence, $\hat{c}_m$ is the $m$-th element of the predicted muscle conditions $\hat{C}_\mathrm{muscle}$, and $\mathrm{D}_\mathrm{g}$ is the difference between two gait patterns. The first term is the reconstruction loss between input and output gaits, and the second term is the typical kl divergence loss for VAE learning, where we use the standard normal distribution $N(0,I)$ as our prior. The last term regularizes the predicted conditions to the reference conditions (i.e., typical healthy adults). 

Note that the probability distribution is located in the middle of the backward model rather than its output $\hat{C}_\mathrm{muscle}$ so that muscle conditions are encoded in the low-dimensional latent space, where the latent codes are then decoded to muscle conditions via the \textit{pre-decoder} (see Figure~\ref{fig:bidirectional_gaitnet}). 
This structure has several advantages over modeling muscle conditions directly as stochastic latent variables.  First, many muscles are functionally correlated, meaning that muscle coordination can have lower intrinsic dimensions and potentially be modeled efficiently in low-dimensional spaces.  Second, the multi-modality of muscle conditions can be better preserved.  In c-VAE, the latent space is often modeled as a uni-modal probability distribution, which is not ideal for our case because there exist disparate muscle conditions that generate almost the same gait.  The \textit{pre-decoder} structure enables us to map a simple uni-modal distribution to a complex multi-model distribution. 

\subsection{Learning}

\subsubsection{Selection of Muscle Parameters}
The musculoskeletal model we use has 604 muscle conditions in total which cover the upper and lower bodies. We observed that prediction of muscle conditions for the upper body (especially for the arms) could harm the overall performance because the upper body motion is loosely correlated with their muscle conditions due to higher freedom in motion when compared to the lower body. Intuitively speaking, people could perform almost any motion while walking, which would not be true for the lower body. To mitigate this problem, we only include muscle conditions for the lower body as well as a few conditions of the muscles attached to the hip (e.g., iliopsoas), of which size is 280 in total. 

\subsubsection{Mixture of Multiple \textit{Backward GaitNet}}
The space of muscle conditions covered by our musculoskeletal model is highly diverse, and we also aim to infer extreme muscle conditions occurred in many patients. To better capture such high variations, we learn multiple \textit{Backward GaitNet}s, then choose one of their output by evaluating the results qualitatively and quantitatively (if possible). This strategy is similar to mixture-of-experts~\cite{masoudnia2014mixture} where each expert covers a subspace of the entire space. Furthermore, this process is analogous to diagnosing a single patient by several doctors who are good at diagnosing a specific pathology. In this work, we use three different models. One is trained only for the muscle conditions of knees and ankles, the other two are trained for the entire muscle with different weights in the Equation~\ref{eq:loss:BG}.

\subsubsection{Data Sampling}
In this work, we are dealing with very high-dimensional space to explore, spanned by 280 conditioning variables. Each conditioning variable has its valid range specified by the user. In order to examine only the corners of this high-dimensional domain, we need to sample a tremendously large number $2^{280} \approx 1.9 \times 10^{84}$ of condition-gait tuples, which is computationally intractable. Surprisingly, the training of both \textit{Forward and Backward GaitNet} required a much smaller collection (1.7 million) of condition-gait tuples in our experiments. Both networks generalize pretty well to learn the influence of physical conditions on gait from very sparse samples. Admitting that sparse sampling is inevitable, uniformly random sampling in the high-dimensional space is not ideal for exploring large variations in physical conditions. We use an alternative approach called grid-based sampling, which selects samples only at corner points, where all conditions can have either minimum or maximum value in their valid range. In our experiments, grid-based sampling outperforms uniform random sampling in particular for reproducing severely impaired gaits. The rationale for our grid-based sampling is that changes in motion and muscle conditions often have monotonically increasing/decreasing relationship.  For example, interpolation of the two different pathological gaits, where one is generated by lengthening a specific muscle and the other is vice-versa, can often produce a gait that can be generated when the muscle is in its normal condition. This implies that having the model experience the extreme input conditions by combining those extremities would be more efficient than uniform sampling in the specified range.

\section{Results}

To train \textit{Bidirectional GaitNet}, we collected approximately 500 hours long walking motions from our predictive gait simulator. This data generation takes approximately 10 hours in the cluster machine equipped with 20 Intel Xeon 6242 CPUs. We use \textit{Torch}~\cite{paszke2019pytorch} to implement our models. The encoder, the pre-decoder, and \textit{Forward GaitNet} are modeled by feed-forward networks with [256, 256, 256], [256, 256, 256], and [512, 512, 512] layers, respectively. For the activation units, LeakyReLu/Linear, LeakyReLu/Sigmoid, and ReLu/Linear are used for the hidden/output layers. 
Both models are learned by following the standard procedure of supervised learning with 65536 (forward), 2048 (backward) batch sizes, \textit{Adam} optimizer with $1e-5$ learning rate. The training takes 2, 1.5 hours for \textit{Forward GaitNet}, and \textit{Backward GaitNet}, respectively, by using the desktop equipped with Ryzen 3950, NVIDIA 2070, and 64GB ram. The dimension of $M$, $z$, $(C_\mathrm{gait}, C_\mathrm{skeleton})$, and $C_\mathrm{muscle}$ are 6060, 32, 13, and 268, respectively.

\subsection{Evaluation on Unseen Simulated Data} \label{sec:results:evaluation_simulated}

Our \textit{Bidirectional GaitNet} is trained by using the dataset generated by a predictive gait simulator. We first evaluate that our model is generalizable to 51 unseen simulated data points separated out from the training data, which includes normal, fully random, and pathological cases. Out of 51 data points, 7 data points are created by running \textit{Generative GaitNet} with the similar muscle conditions demonstrated in the previous work [Park et al. 2022] to cover well-known gait patterns (normal, foot drop, equinus, stiff knee, crouch, trendelenburg, and waddling gait) while the remaining data points are created by randomly sampling anatomical conditions within the valid range. Note that they are generated independently, so exactly the same data are never included in the training dataset.

\subsubsection{Prediction of Gaits}
We ultimately should be able to predict anatomical conditions that are realizable similarly to the input gait in the predictive gait simulator. So, we measure error between the ground truth gait and the simulated gait with the predicted anatomical conditions from \textit{Backward GaitNet}, where the 3rd row in Table~\ref{table:joint_prediction_error} shows the average and variance of joint angle prediction error measured over 2 gait cycles. On average, the angle difference is less than 8 degrees. This means that our model is able to generalize to unseen simulated data successfully. Furthermore, the amount of error occurred by our model is almost visually indistinguishable as demonstrated in both Figure~\ref{fig:gait_prediction} and the supplemental video.

\begin{table*}[]
\caption{\label{table:joint_prediction_error} Joint Angle Prediction Error.}
\resizebox{\textwidth}{!}{%
\begin{tabular}{@{}c|c|cccccccccccc@{}}
\toprule
\begin{tabular}[c]{@{}c@{}}Forward\\ GaitNet\end{tabular} & \begin{tabular}[c]{@{}c@{}}BackWard\\ GaitNet\end{tabular} & Pelvis                                                                & FemurR                                                                 & TibiaR                                                                & TalusR                                                               & FootPinkyR                                                            & FootThumbR                                                            & FemurL                                                                & TibiaL                                                               & TalusL                                                                & FootPinkyL                                                            & FootThumbL                                                            &           \\ \midrule
Uniform                                                   & Uniform                                                    & \begin{tabular}[c]{@{}c@{}}8.87398\\ (2.27329)\end{tabular}           & \begin{tabular}[c]{@{}c@{}}11.9979 \\ (1.97183)\end{tabular}           & \textbf{\begin{tabular}[c]{@{}c@{}}5.53967 \\ (6.22365)\end{tabular}} & \begin{tabular}[c]{@{}c@{}}14.874 \\ (7.87755)\end{tabular}          & \begin{tabular}[c]{@{}c@{}}8.38405 \\ (8.08005)\end{tabular}          & \begin{tabular}[c]{@{}c@{}}8.89714 \\ (8.29367)\end{tabular}          & \begin{tabular}[c]{@{}c@{}}12.2464 \\ (7.78203)\end{tabular}          & \textbf{\begin{tabular}[c]{@{}c@{}}5.33964 \\ (2.2433)\end{tabular}} & \begin{tabular}[c]{@{}c@{}}14.6996 \\ (2.27785)\end{tabular}          & \begin{tabular}[c]{@{}c@{}}7.49264 \\ (8.47628)\end{tabular}          & \begin{tabular}[c]{@{}c@{}}7.35133 \\ (12.6308)\end{tabular}          &           \\
Uniform                                                   & Grid                                                       & \begin{tabular}[c]{@{}c@{}}8.77184 \\ (4.27071)\end{tabular}          & \textbf{\begin{tabular}[c]{@{}c@{}}11.8603 \\ (0.746043)\end{tabular}} & \begin{tabular}[c]{@{}c@{}}5.47452 \\ (5.75071)\end{tabular}          & \begin{tabular}[c]{@{}c@{}}15.085 \\ (4.78839)\end{tabular}          & \begin{tabular}[c]{@{}c@{}}8.38307 \\ (8.30824)\end{tabular}          & \begin{tabular}[c]{@{}c@{}}8.45141 \\ (8.17749)\end{tabular}          & \begin{tabular}[c]{@{}c@{}}12.2202 \\ (5.44729)\end{tabular}          & \begin{tabular}[c]{@{}c@{}}5.3674 \\ (5.05303)\end{tabular}          & \begin{tabular}[c]{@{}c@{}}14.6827 \\ (7.69895)\end{tabular}          & \textbf{\begin{tabular}[c]{@{}c@{}}7.29844 \\ (12.5078)\end{tabular}} & \textbf{\begin{tabular}[c]{@{}c@{}}7.00184 \\ (1.70778)\end{tabular}} &           \\
Grid                                                      & Grid                                                       & \textbf{\begin{tabular}[c]{@{}c@{}}8.47702 \\ (2.15696)\end{tabular}} & \begin{tabular}[c]{@{}c@{}}11.8692 \\ (1.15372)\end{tabular}           & \begin{tabular}[c]{@{}c@{}}5.79214 \\ (11.8167)\end{tabular}          & \textbf{\begin{tabular}[c]{@{}c@{}}14.7716 \\ (2.1861)\end{tabular}} & \textbf{\begin{tabular}[c]{@{}c@{}}7.62802 \\ (6.71003)\end{tabular}} & \textbf{\begin{tabular}[c]{@{}c@{}}7.67742 \\ (7.66728)\end{tabular}} & \textbf{\begin{tabular}[c]{@{}c@{}}11.1595 \\ (1.70138)\end{tabular}} & \begin{tabular}[c]{@{}c@{}}5.4935 \\ (1.85809)\end{tabular}          & \textbf{\begin{tabular}[c]{@{}c@{}}14.5509 \\ (1.28581)\end{tabular}} & \begin{tabular}[c]{@{}c@{}}8.45036 \\ (20.0347)\end{tabular}          & \begin{tabular}[c]{@{}c@{}}8.14378 \\ (22.3314)\end{tabular}          & \textbf{} \\ \bottomrule
\end{tabular}%
}


\resizebox{\textwidth}{!}{%
\begin{tabular}{@{}cccccccccccc@{}}
\toprule
Spine                                                                 & Torso                                                                  & Neck                                                                  & Head                                                                 & ShoulderR                                                              & ArmR                                                                  & ForeArmR                                                              & HandR                                                                & ShoulderL                                                             & ArmL                                                                  & ForeArmL                                                              & HandL                                                                 \\ \midrule
\begin{tabular}[c]{@{}c@{}}8.59933 \\ (1.50475)\end{tabular}          & \begin{tabular}[c]{@{}c@{}}8.22234 \\ (1.96971)\end{tabular}           & \begin{tabular}[c]{@{}c@{}}7.4927 \\ (4.40804)\end{tabular}           & \textbf{\begin{tabular}[c]{@{}c@{}}6.8291 \\ (4.05152)\end{tabular}} & \begin{tabular}[c]{@{}c@{}}1.80658 \\ (0.316834)\end{tabular}          & \begin{tabular}[c]{@{}c@{}}11.5427 \\ (6.8967)\end{tabular}           & \begin{tabular}[c]{@{}c@{}}3.24304 \\ (0.75264)\end{tabular}          & \textbf{\begin{tabular}[c]{@{}c@{}}9.78388 \\ (7.1614)\end{tabular}} & \begin{tabular}[c]{@{}c@{}}1.98267 \\ (1.95443)\end{tabular}          & \begin{tabular}[c]{@{}c@{}}10.9824 \\ (1.56775)\end{tabular}          & \begin{tabular}[c]{@{}c@{}}3.00934 \\ (2.98694)\end{tabular}          & \begin{tabular}[c]{@{}c@{}}8.65188 \\ (8.31125)\end{tabular}          \\
\begin{tabular}[c]{@{}c@{}}8.67501 \\ (3.51157)\end{tabular}          & \begin{tabular}[c]{@{}c@{}}8.27154 \\ (0.26088)\end{tabular}           & \begin{tabular}[c]{@{}c@{}}7.57086 \\ (4.38966)\end{tabular}          & \begin{tabular}[c]{@{}c@{}}6.83435 \\ (4.60315)\end{tabular}         & \begin{tabular}[c]{@{}c@{}}1.79577 \\ (0.985541)\end{tabular}          & \begin{tabular}[c]{@{}c@{}}11.52 \\ (4.41658)\end{tabular}            & \begin{tabular}[c]{@{}c@{}}3.25868 \\ (0.643535)\end{tabular}         & \begin{tabular}[c]{@{}c@{}}9.98208 \\ (7.61925)\end{tabular}         & \begin{tabular}[c]{@{}c@{}}1.99904 \\ (1.85795)\end{tabular}          & \begin{tabular}[c]{@{}c@{}}10.9108 \\ (0.524045)\end{tabular}         & \begin{tabular}[c]{@{}c@{}}3.09474 \\ (3.09089)\end{tabular}          & \begin{tabular}[c]{@{}c@{}}8.7093 \\ (5.95637)\end{tabular}           \\
\textbf{\begin{tabular}[c]{@{}c@{}}8.28418 \\ (4.57552)\end{tabular}} & \textbf{\begin{tabular}[c]{@{}c@{}}8.14461 \\ (0.243296)\end{tabular}} & \textbf{\begin{tabular}[c]{@{}c@{}}7.24122 \\ (0.91435)\end{tabular}} & \begin{tabular}[c]{@{}c@{}}6.87393 \\ (3.44239)\end{tabular}         & \textbf{\begin{tabular}[c]{@{}c@{}}1.77189 \\ (0.147876)\end{tabular}} & \textbf{\begin{tabular}[c]{@{}c@{}}10.8449 \\ (1.71357)\end{tabular}} & \textbf{\begin{tabular}[c]{@{}c@{}}2.98214 \\ (1.62909)\end{tabular}} & \begin{tabular}[c]{@{}c@{}}10.2911 \\ (6.87899)\end{tabular}         & \textbf{\begin{tabular}[c]{@{}c@{}}1.92183 \\ (1.64011)\end{tabular}} & \textbf{\begin{tabular}[c]{@{}c@{}}10.5057 \\ (1.13507)\end{tabular}} & \textbf{\begin{tabular}[c]{@{}c@{}}2.74626 \\ (2.73175)\end{tabular}} & \textbf{\begin{tabular}[c]{@{}c@{}}8.04477 \\ (6.98929)\end{tabular}} \\ \bottomrule
\end{tabular}
}

\end{table*}

\subsubsection{Prediction of Muscle Conditions}
\begin{figure}
    \centering
    \includegraphics[width=0.99\linewidth]{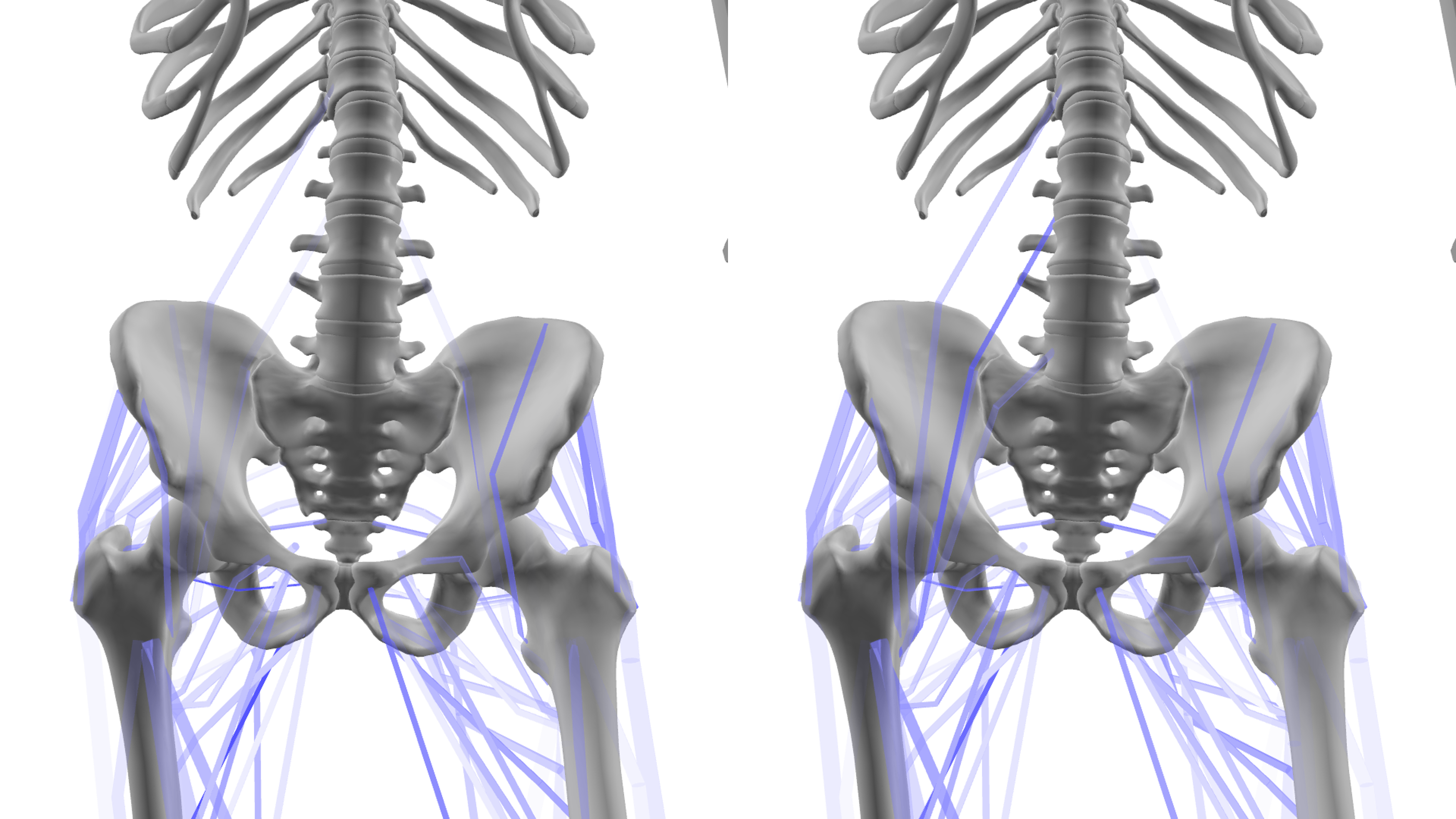}
    \caption{Different muscle conditions given an input gait (a.k.a. trendelenburg).}
    \label{fig:multiple_muscle_conditions}
\end{figure}

Figure~\ref{fig:multiple_muscle_conditions} shows muscle conditions predicted by our \textit{Backward GaitNet} given a typical pathological gait, hip-drop (trendelenburg), which result from defective muscles at one side of hip. Physiologically, such dropping motion could appear either when a muscle at one side has contracture or when the other side has weakness. Our results show that both physiologically plausible conditions can be discovered successfully by our backward model. In the supplemental video, we also show a variety of predicted muscle conditions and their corresponding simulated gait.

\begin{figure}
    \centering
    \begin{subfigure}{0.99\linewidth}
        \includegraphics[width=0.49\linewidth]{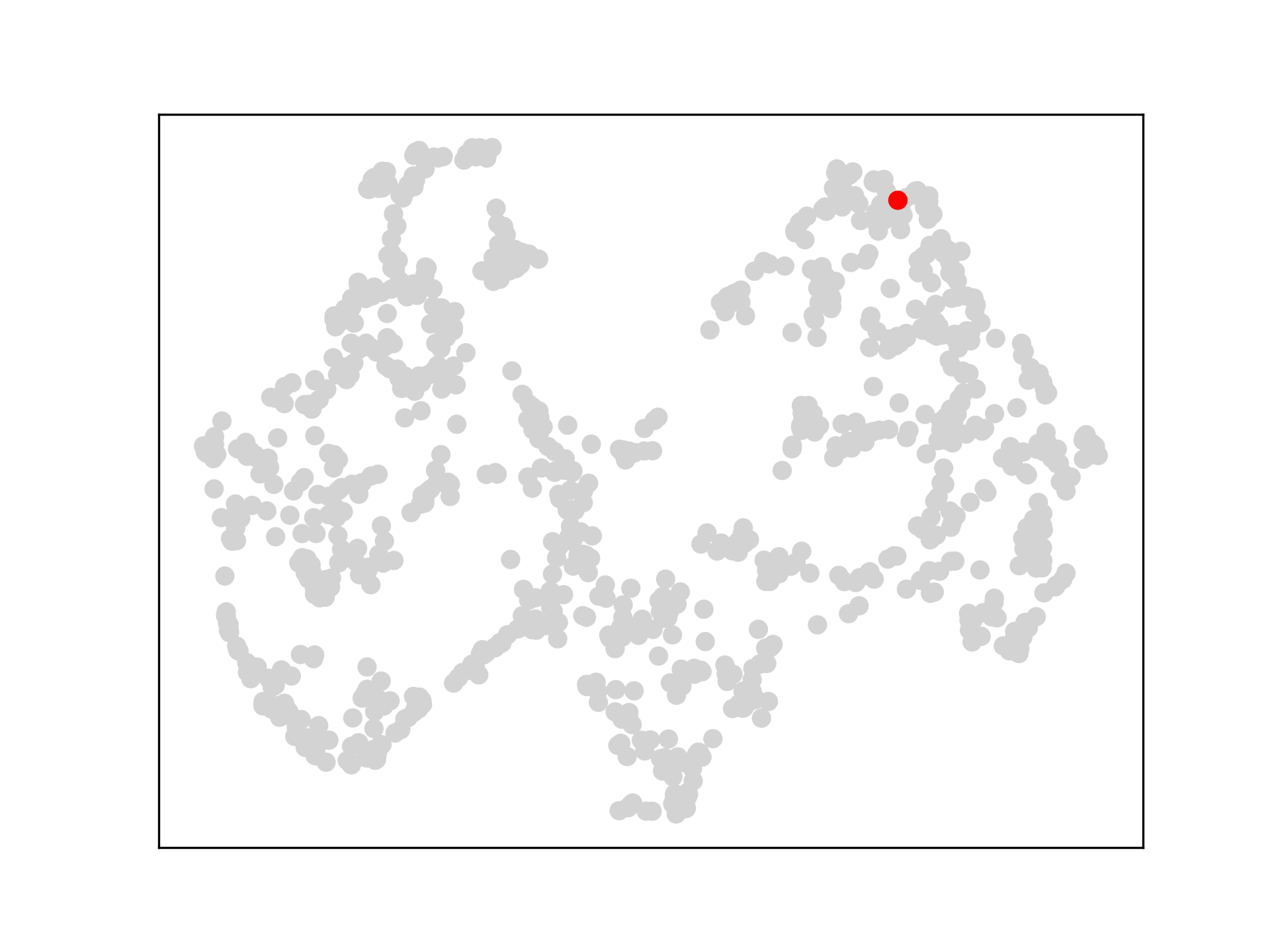}
        \includegraphics[width=0.49\linewidth]{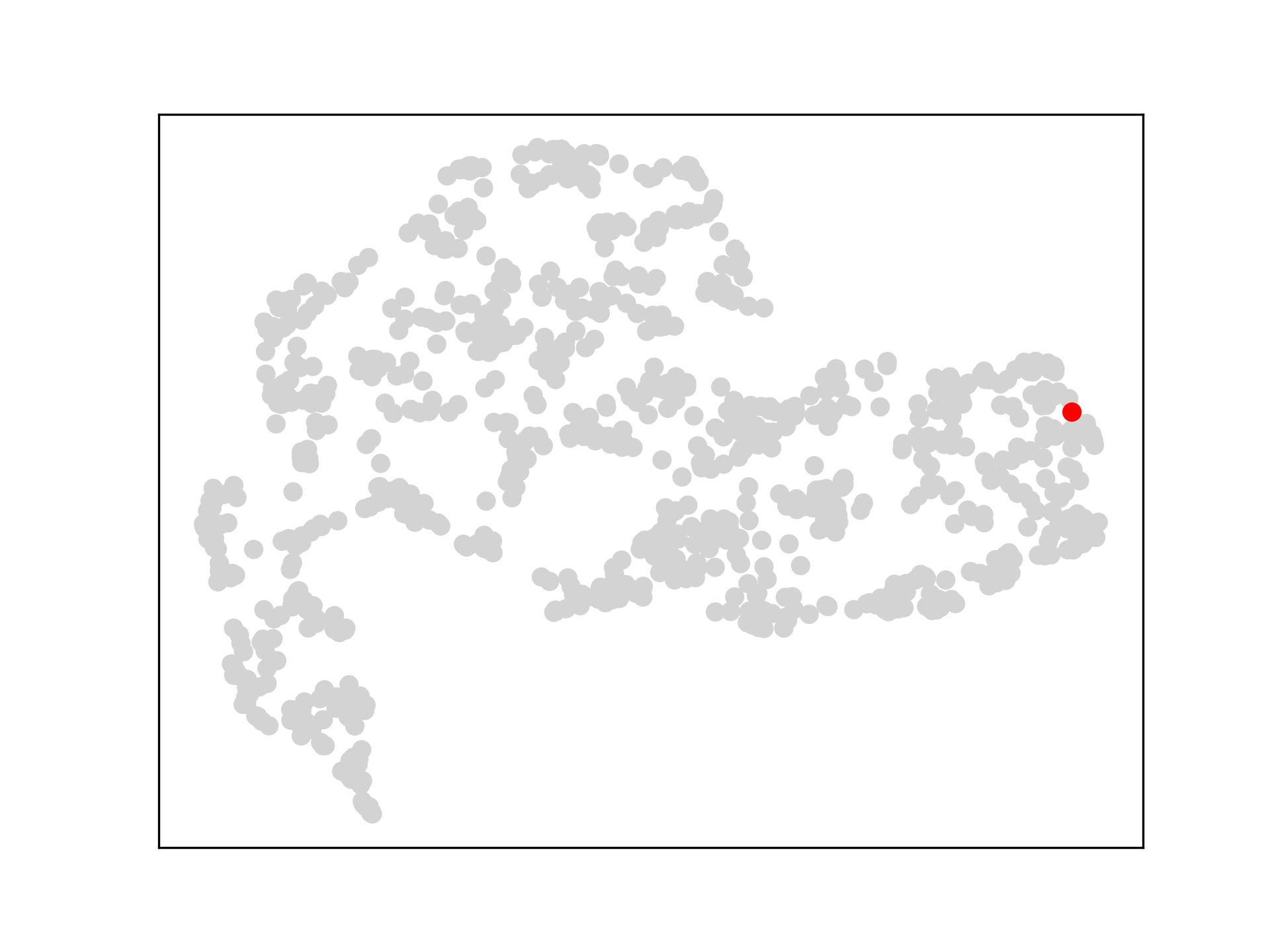}
    \end{subfigure}
    \hfill
    \begin{subfigure}{0.99\linewidth}
        \includegraphics[width=0.49\linewidth]{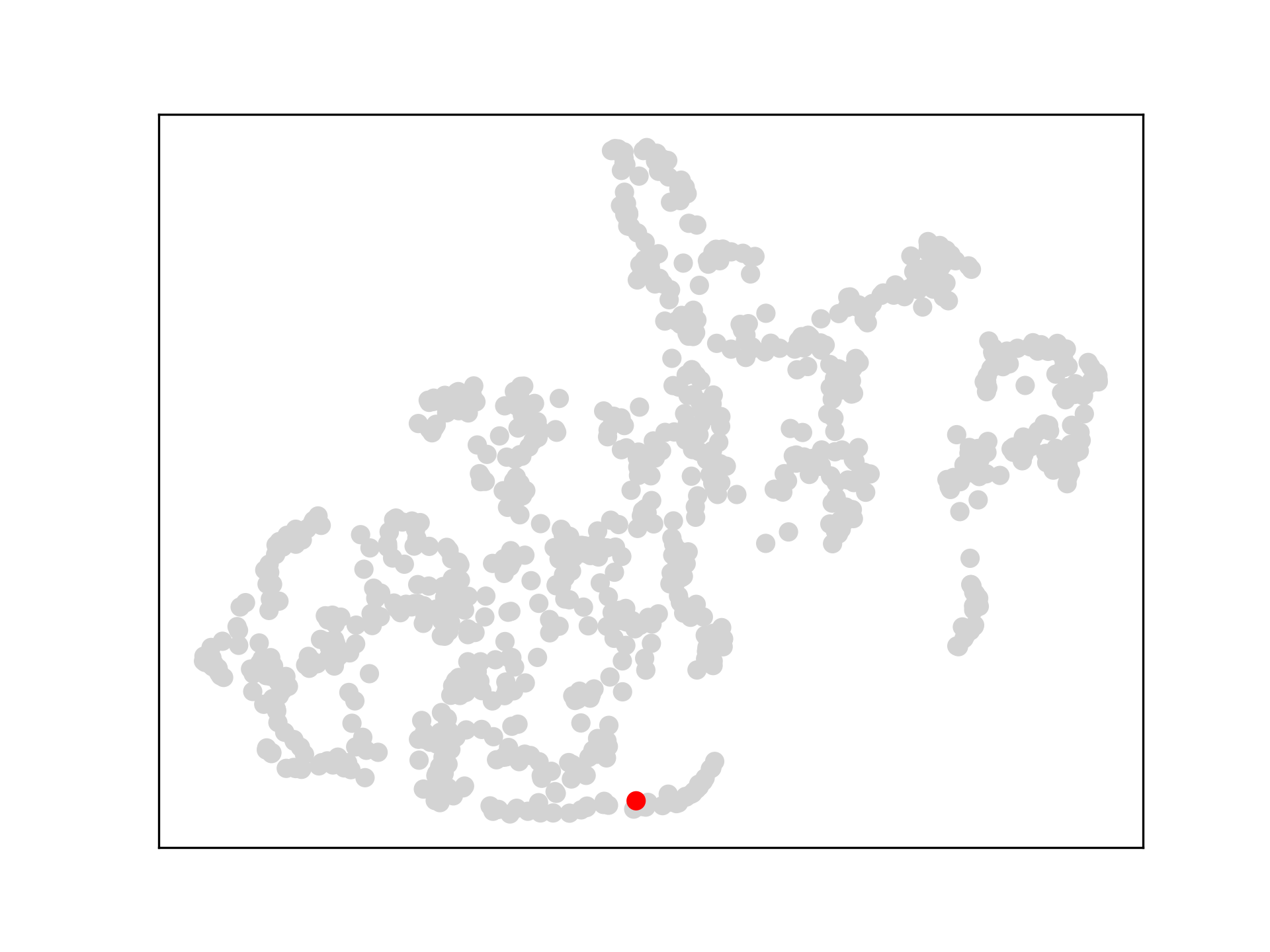}
        \includegraphics[width=0.49\linewidth]{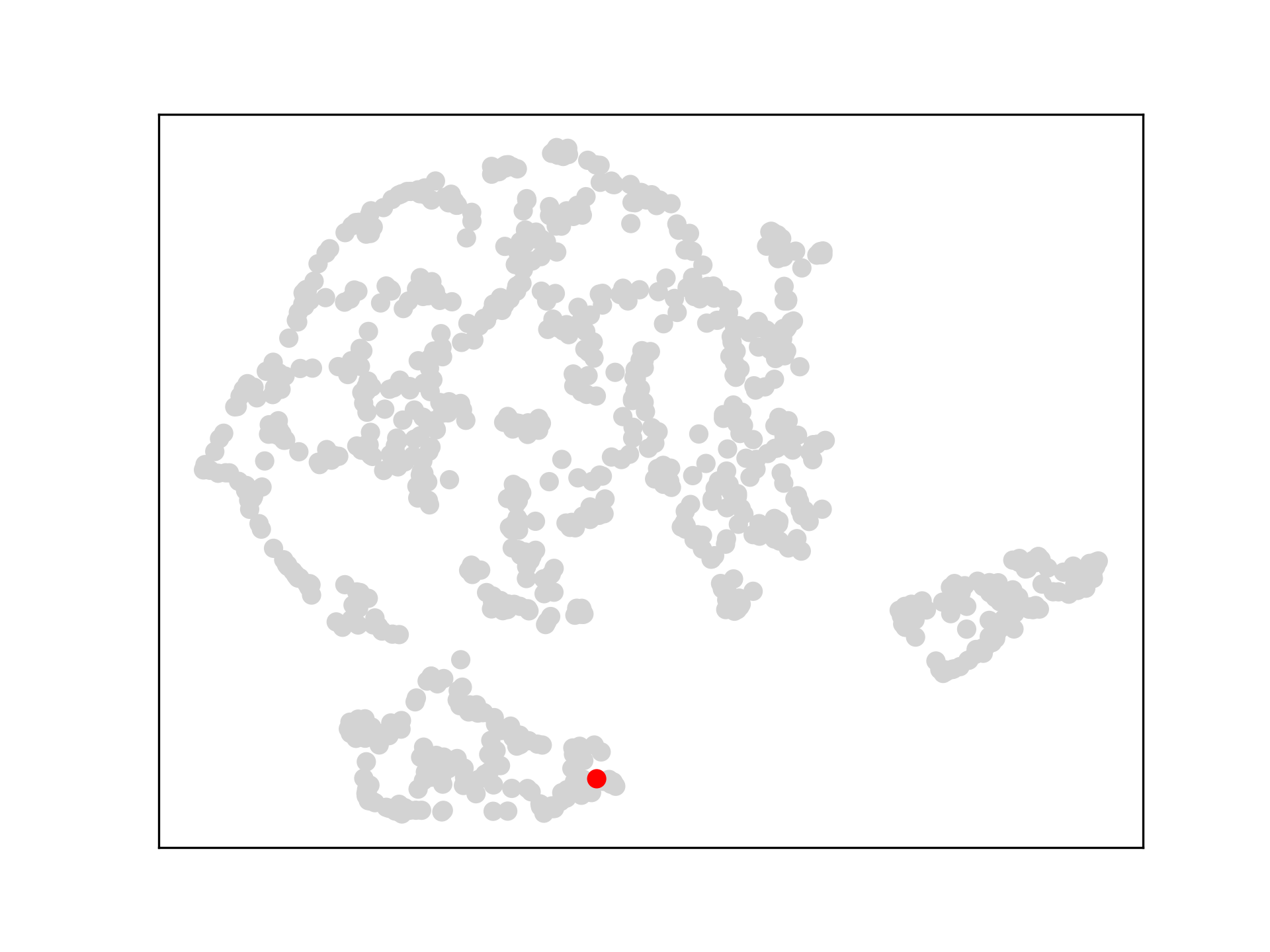}
    \end{subfigure}
    \caption{2D Embeddings of predicted and ground-truth muscle conditions over 4 different input gaits. The grey and red dots represent predicted and ground-truth conditions, respectively, the ground truth conditions are observable with high probability under the distribution of predicted muscle conditions.}
    \label{fig:umap_comparison}
\end{figure}

We also compare the ground truth muscle conditions with the estimated muscle conditions.  In contrast to gait prediction error, where we can compute joint angle difference directly, the ground truth and estimated muscle conditions are not directly comparable because multiple conditions that generate the same gait pattern could exist.  Instead, given an input gait paired with the ground truth muscle conditions, we evaluate our backward model by testing whether the probability distribution predicted by our model from the input gait includes the ground truth conditions or not.  We first create 1000 sets of muscle conditions from our backward model by randomly sampling from the distribution, then we run a dimensionality technique for those conditions in addition to the ground truth conditions.  Figure~\ref{fig:umap_comparison}(a) shows 2D embedding drawn by Umap~\cite{mcinnes2018umap}, from which we could infer that the ground truth conditions are observable with high probability under the predicted distribution.

\subsection{Evaluation on Gaits of Real Patients}
\begin{figure*}
    \centering
    \includegraphics[width=0.9\linewidth]{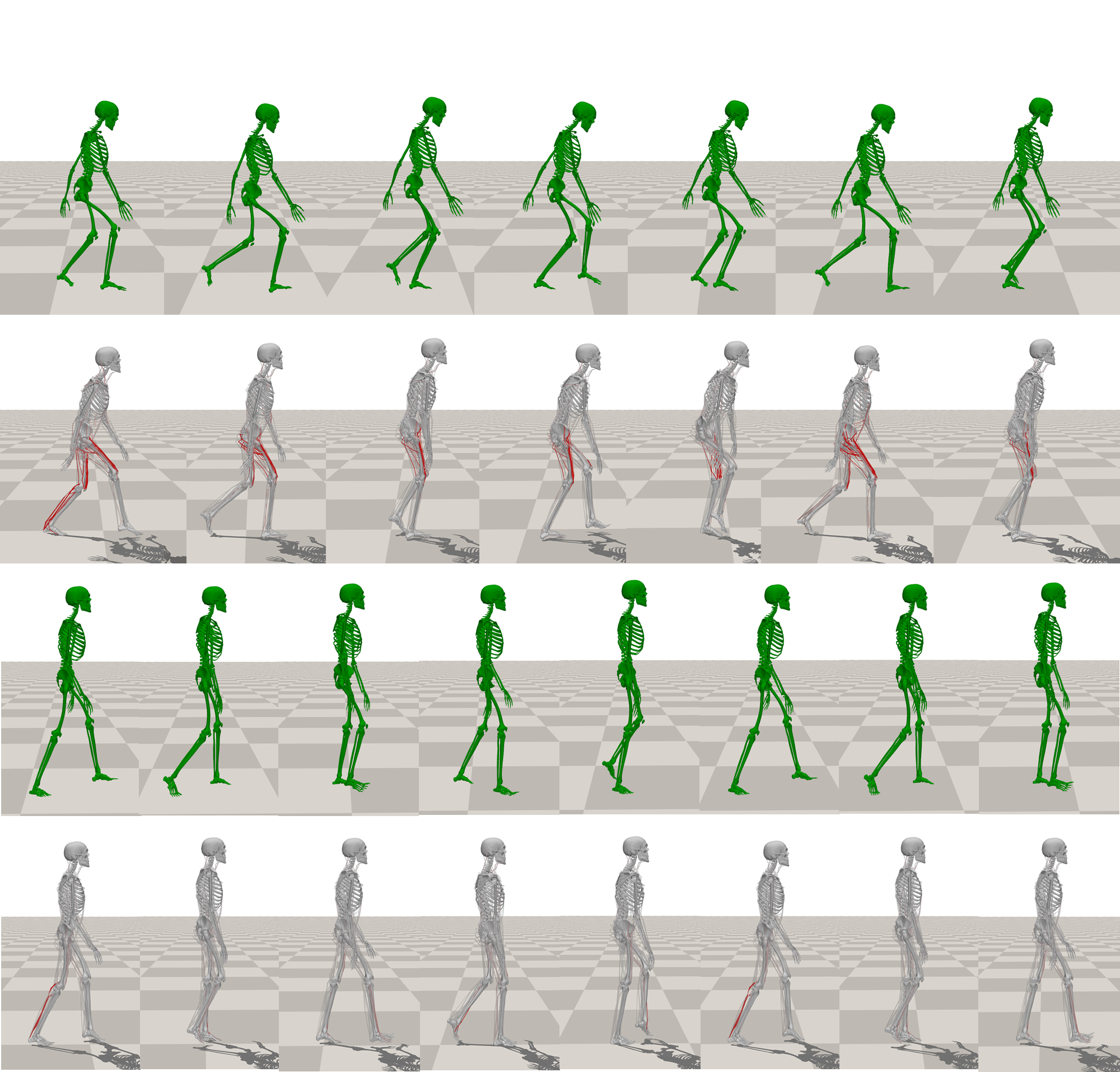}
    \caption{A comparison of input gaits (green) and their simulated gaits (white) with predicted anatomical conditions from our \textit{Backward GaitNet}.}
    \label{fig:gait_prediction}
\end{figure*}

We show further generalization capability by evaluating our model for gaits of real patients.  Unlike the evaluation on simulated data, the ground truth muscle conditions do not exist for real patients because examining such conditions for every muscle of a patient is not realizable rather examining itself is a very challenging problem.  We evaluate our results qualitatively by comparing the input gaits from real patients and the simulated gaits under the anatomical conditions predicted by our backward model (see Figure~\ref{fig:gait_prediction} and our supplemental video).  Even if we trained our model with the simulation data only, our model still can produce plausible predictions for the gaits from real patients which might differ from simulation results due to \textit{Sim-to-Real} gap.

\begin{figure}
    \centering
    \begin{subfigure}{0.9\linewidth}
        \includegraphics[width=0.49\linewidth,trim=3 0 0 0,clip]{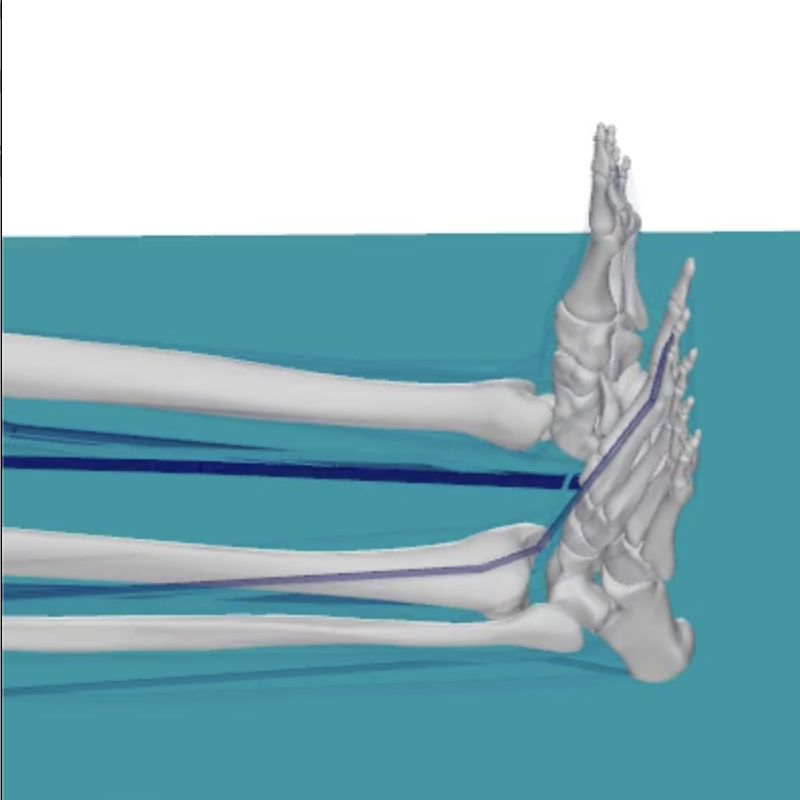}
        \includegraphics[width=0.49\linewidth,trim=3 0 0 0,clip]{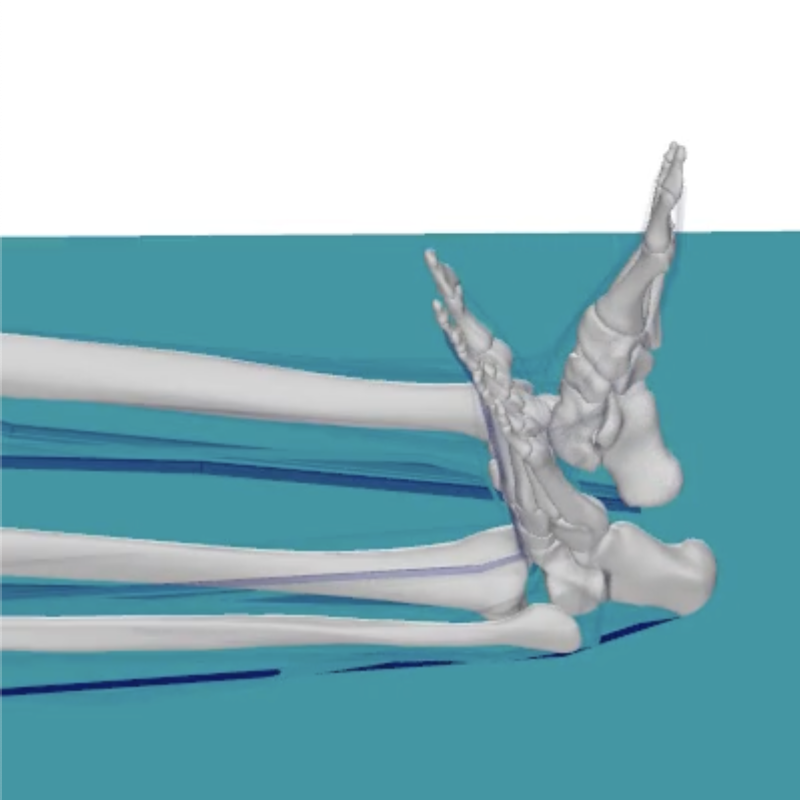}
    \end{subfigure}
    \caption{Physical Exam on Ankles}
    \label{fig:evaluation:physical_exam}
\end{figure}

We also evaluate the predicted muscle conditions by using \textit{Physical Exam}~\cite{moon2017normative}, which has been widely used in medicine for testing muscle conditions of patients.  The exam measures range-of-motions of several joints at pre-specified postures to examine that muscle lengths are either shortened or lengthened permanently, which significantly could affect functionality in locomotion if they differ from the normal conditions.  We implemented the exam, where the maximum range-of-motion for a joint is determined by measuring the magnitude of joint torque accumulated from passive forces of muscle relevant to the joint while performing the exam, where we regard it as the maximum value if the magnitude is larger than 20 Nm. Figure~\ref{fig:evaluation:physical_exam} compares range-of-motions of the left and the right ankles given a patient's gait, where our system predicts that the reason of asymmetry of the input gait is because the right calf muscles are weaker than the left one. This study is approved by The Institutional Review Board of Seoul National University Hospital (B-1107-132-101).

\subsection{Comparison}

To show the effectiveness of the technical components adopted in our method, we conduct ablation studies on the three components that we think are crucial.

\subsubsection{Grid vs. Uniform Sampling}
\label{sec:comparison:sampling}

We show the effectiveness of grid-based random sampling which we used when generating the training data, by comparing it with another model trained with data generated by the uniform sampling.  We run an evaluation with the same unseen simulated gaits used in Section~\ref{sec:results:evaluation_simulated}. Table~\ref{table:joint_prediction_error} compares joint angle prediction errors over three different design choices, \textit{Uniform-Uniform}, \textit{Uniform-Grid}, and \textit{Grid-Grid} (ours), where the errors are lower in general when our grid-based sampling is used for learning both the forward and the backward models.

\section{Discussion}

We developed a novel generative model, called \textit{Bidirectional GaitNet}, that learns the relationship between human anatomy and its gait. It consists of forward and backward models where the forward model predicts a gait pattern of a person with specific physical conditions, while the backward model estimates the physical conditions of a person when his/her gait pattern is provided. By constructing a c-VAE structure with the forward model learned by simulation data generated from the state-of-the-art predictive gait simulator, we were able to learn its inverse mapping (i.e., the backward model) effectively. We showed that the anatomical conditions predicted by our model were able to be realized for both unseen simulated gaits and real patients' gaits.

There are still several limitations in our method.  First, although joint angle prediction error is pretty accurate (less than 8 degrees on average), it needs to be further improved for some joints such as talus (ankle) and femur (upper leg) which show larger errors when compared to other joints.  Note that those are actually the joints that have much wider variations in our dataset than others. The sophisticated network architectures (e.g., Transformer) or loss functions that can better capture wider variations in data would be helpful in general.  Second, our method can be easily extended so that it includes the upper body muscle conditions. However, the prediction of those muscle conditions might be less reliable when compared to ones in the lower body due to weaker dependency between gaits and their muscle conditions. Simply speaking, the upper body motions have much larger freedom and could be almost arbitrary. In addition, some of anatomical features such as the nervous system, flexible tendon model, and skin are missing in the musculoskeletal model we used in this research, which might affect the prediction of gait and physical condition differently. 

We envision several promising future directions that we want to study further. The input of the current system should be given as mocap data, which might be cumbersome and expensive to obtain.  It would be interesting if we can use a monocular video as input to our system, which could be done by using existing video-to-mocap solutions or other training schemes. Another interesting direction would be exploring more complex full-body or facial musculoskeletal models that include volumetric muscles, which will have a practical impact on animation industries because those models have often been used in commercial films.

\begin{acks}
This study was supported by the New Faculty Startup Fund from Seoul National University, ICT(Institute of Computer Technology) at Seoul National University, and grant no (14-2020-0012) from the SNUBH Research Fund.
\end{acks}

\bibliographystyle{ACM-Reference-Format}
\bibliography{reference}


\end{document}